\documentclass{article}

\usepackage{arxiv}
\usepackage{amsmath,amssymb,amsthm,amscd,verbatim, graphicx, geometry, colortbl}
\usepackage[mathscr]{eucal}
\usepackage{bm}
\usepackage{thmtools}
\usepackage[colorlinks, citecolor= blue ,urlcolor=red]{hyperref}       
\usepackage{url}             
\usepackage{booktabs}  
\usepackage{nicefrac}   
\usepackage{microtype}  
\usepackage{doi}
\usepackage{thm-restate}
\usepackage{appendix}
\usepackage{textcomp}
\usepackage{booktabs}
\usepackage{multirow}
\usepackage{xcolor}
\usepackage{epstopdf}
\usepackage{textgreek}
\usepackage{natbib}[authoryear]

\newtheorem{lemma}{Lemma}
\newtheorem{result}{Result}

\theoremstyle{remark}
\newtheorem{example}{Example}

\newcommand*\xbar[1]{%
   \hbox{%
     \vbox{%
       \hrule height 0.5pt 
       \kern 0.5ex
       \hbox{%
         \kern-0.18em
         \ensuremath{ #1 }%
         \kern-0.1em
       }%
     }%
   }%
}

\newcommand{\IF}{\text{IF}}
\newcommand{\HIF}{\text{HIF}}
\newcommand{\SIF}{\text{SIF}}
\newcommand{\EIF}{\text{EIF}}
\newcommand{\Fe}{F_\epsilon}
\newcommand{\e}{\epsilon}
\newcommand{\X}{\bm{X}}

\newcommand{\Si}{\bm{\Sigma}}
\newcommand{\g}{\bm{\eta}}

\newcommand{\x}{\mathbf{x}_0}

\newcommand{\norm}[1]{\left\lVert#1\right\rVert}

\title{A note on switching eigenvalues under small perturbations}

\author{ 
{Marina Masioti}\\
	Department of Mathematical and Physical Sciences\\
	La Trobe University\\
	Melbourne, VIC 3086, Australia\\
	\texttt{mmasioti@students.ltu.edu.au} \\
	\And
	Connie S. N. Li-Wai-Suen \\
	Walter and Eliza Hall Institute of Medical Research \\
	\texttt{liwaisuen@wehi.edu.au} \\
	\AND
	\href{https://scholars.latrobe.edu.au/lprendergast}{\includegraphics[scale=0.06]{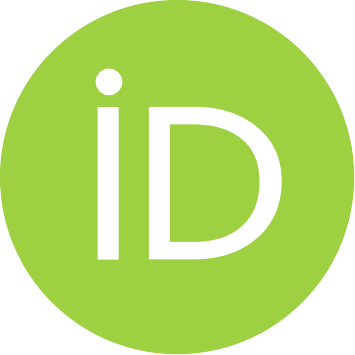}\hspace{1mm}Luke A. Prendergast} \\
    Department of Mathematical and Physical Sciences \\ 
    La Trobe University,
	Melbourne, Australia \\
	\texttt{luke.prendergast@latrobe.edu.au} \\
	\And
	Amanda Shaker\\
	Department of Mathematical and Physical Sciences\\ 
	La Trobe University, Melbourne, Australia \\
	\texttt{A.Shaker@latrobe.edu.au} \\
}

\renewcommand{\shorttitle}{Switching eigenvalues}

\hypersetup{
pdftitle={OTDR},
pdfsubject={q-bio.NC, q-bio.QM},
pdfauthor={Marina Masioti, Connie Li Wai Suen, Luke A. Prendergast, Amanda Shaker},
pdfkeywords={Principal component analysis, switching, eigenvalues},
}

\begin{document}
\maketitle

\begin{abstract}
Sensitivity of eigenvectors and eigenvalues of symmetric matrix estimates to the removal of a single observation have been well documented in the literature.  However, a complicating factor can exist in that the rank of the eigenvalues may change due to the removal of an observation, and with that so too does the perceived importance of the corresponding eigenvector.  We refer to this problem as ``switching of eigenvalues".  Since there is not enough information in the new eigenvalues post observation removal to indicate that this has happened, how do we know that this switching has occurred?  In this paper we show that approximations to the eigenvalues can be used to help determine when switching may have occurred. We then discuss possible actions researchers can take based on this knowledge, for example making better choices when it comes to deciding how many principal components should be retained and adjustments to approximate influence diagnostics that perform poorly when switching has occurred. Our results are easily applied to any eigenvalue problem involving symmetric matrix estimators.  We highlight our approach with application to a real data example. 
\end{abstract}


\section{Introduction}

In the past 30 years or so, statisticians have been particularly concerned with developing and improving methods for the analysis of high-dimensional data. That is, multivariate data sets that contain a large number of measurement variables. Dimension reduction was formulated as a high-dimensional data visualization problem, where a new set of low-dimensional projected variables are found that can be used to explore relationships between variables.  Our interest is in methods that require an eigen-decomposition of a symmetric matrix.  The most common of these is Principal Component Analysis \citep[PCA,][]{Pearson1901} which eigen-decomposes the covariance or correlation matrix.  However, other methods like Sliced Inverse Regression \citep[SIR,][]{Li91} can be useful when a response variable is available and these also require an eigen-decomposition of a symmetric matrix in order to obtain the lower-dimensional projected variables.


Many estimators can be very sensitive to outliers and other types of highly influential observations which can be detrimental to the results of the analysis. Therefore, it is very important to detect observations that can greatly affect the estimation and to develop simple techniques to deal with them. The Influence Function \citep[IF]{hampel1974influence}, and other influence diagnostics that have been developed for the dimension reduction setting, have been used by many authors to study the sensitivity of estimators of interest and to detect influential observations  \citep[see, for example,][]{ critchley1985influence, Benasseni1990SensitivityCF, croux2000principal, PRENDERGAST_2005, Pren06, Smith2007, Smith10}.

In practice, sample influence diagnostics utilize the leave-one-out approach to find the influence of each observation on an estimator, by considering the removal of the observation from the data and comparing the estimator pre- and post- removal. In some cases, removing a single observation from the estimation can affect the rank of the eigenvalues, so that the order of two consecutive eigenvalues might change, which in turn affects the corresponding eigenvectors. This problem is referred to as ``switching of eigenvalues" (or just ``switching", for simplicity), since it disrupts the usual decreasing order of the eigenvalues. Dimension reduction methods can be highly sensitive to such observational types. 

The problem of switching has been briefly mentioned by \cite{critchley1985influence} in the context of PCA. Much later, second-order approximations to the eigenvalues of PCA were used to detect switching \citep{liwaisuenthesis}. Our goal in this paper is to consider approximations to the eigenvalues that enable switching to be detected for methods that involve the eigen-decomposition of a symmetric matrix. We also provide a few recommendations of possible actions one can take to improve the results of the analysis when switching occurs at critical places. Throughout, we will present and explore switching in the context of PCA.

We introduce PCA and related influence diagnostics in Section \ref{sec:PCA}. We then demonstrate the problem of switching through a motivating example in Section \ref{sec:Switch} and provide simple approximations to the eigenvalues, applicable to any symmetric matrix. In Section \ref{sec:Examples}, a real-world example is used to highlight the impact of switching in practice. Recommendations on possible actions for when switching occurs are given in Section \ref{sec:Suggestions}. Finally, a discussion and future work are considered in Section \ref{sec:Discuss}.

\section{Principal Component Analysis}\label{sec:PCA}

The aim of dimension reduction methods, like PCA, is to explain most of the variability in high-dimensional data in just a few summary measures, without much loss of information. These summary measures are linear combinations of the original variables and form a new set of uncorrelated variables, which are significantly fewer than the former. PCA is especially useful in many areas of research due to its applicability on data sets where the number of measurement variables ($p$) exceeds the number of observations ($n$) such that $n < p$, or $n << p$, for example in RNA-seq data. 

For a $p$-dimensional random vector $\X$, PCA is performed through an eigen-decomposition on the covariance matrix of $\X$, denoted by $\Si$, which is assumed to be positive semi-definite. When the variables in $\X$ have different unit measurements then the correlation matrix is used instead.  While we focus on the covariance matrix where $\g_j^\top \Si = \lambda_j \g_j^\top$ (for $j = 1 ,\dots, p$), the suggestions that follow are easily adapted to PCA with the correlation matrix and many other estimators. Denote the eigenvalues $\lambda_1, \dots, \lambda_p$ and their corresponding $p$-dimensional orthonormal eigenvectors $\g_1, \dots, \g_p$, such that $\lambda_1 \geq \lambda_2 \geq \dots \geq \lambda_p \geq 0$.  Then, the new set of variables, called principal components (PCs) is given by, $\alpha_j = \g_j^\top \X$, for $j = 1, \dots, p$.  In the sample setting, $\widehat{\alpha}_j = \widehat{\g}_j^\top \X$ is the $j^{th}$ sample principal component (SPC), obtained from the eigen-decomposition of the sample covariance matrix, $\widehat{\Si}$.

Further, var($\bm{\alpha}_j$) = $\lambda_j$, meaning that the $j^{th}$ eigenvalue reveals the proportion of the total variability explained by the $j^{th}$ PC. This provides one way with which we choose the number $L$ of PCs to retain, where a common criterion is to obtain the first $L$ PCs that explain most of the variation in the data, such that $L << p$. Another way is through the scree plot of the eigenvalues, where the first $L$ PCs are the ones up to and including the eigenvalue where the biggest difference in slope is observed. In practice, it is common to choose $L$ to be 1, 2 or 3, to allow for simple visualizations of the data. A detailed discussion of the various ways for choosing $L$ can be found in Chapter 6 of \cite{jolliffe2002principal}.

A notable challenge in PCA is its susceptibility to outliers and influential observations. Sometimes a big enough outlier can itself determine an entire component \citep[see, for example, Exhibit 8.1.1,][]{Huber81}, which can even have one of the largest eigenvalues. While the use of robust estimates of the covariance (or correlation) matrix is one way of dealing with outliers, another would be to remove them from the data. However, the removal of observations might not always be beneficial. Note here that an influential observation is not always an outlier and vice versa. Many authors have studied the sensitivity of PCA using various influence diagnostics for the detection of influential observations \citep[see, for example, ][]{critchley1985influence, Tanaka88, Tanaka90, Benasseni1990SensitivityCF, Prendergast2008, Connie2011}. In this article, we will use influence diagnostics to find accurate approximations to the eigenvalues after leave-one-out perturbations. 

\subsection{Influence diagnostics for PCA}\label{sec:PCA_IFs}

Here we introduce two influence diagnostics that have been developed for PCA. Since we are interested in discussing the influence of observations in the sample setting, we will only provide the sample and empirical influence functions of those diagnostics. Consider a sample of $n$ $p$-dimensional observations denoted $\mathbf{x}_1,\ldots,\mathbf{x}_n$ where $F_n$ is the empirical distribution and $\X_n$ is the $n\times p$ matrix whose $i$th row is $\mathbf{x}_i^\top$. Let $\{\widehat{\lambda}_i, \widehat{\g}_i\}_{i=1}^{p}$ be the eigenvalue-eigenvector pairs of the sample covariance matrix. First we consider a diagnostic for the influence on the estimators of the span of the eigenvectors of interest \citep{Benasseni1990SensitivityCF}. That is, the span of the eigenvectors to be retained. Let $\widehat{\bm{\Gamma}}_{L} = [ \widehat{\g}_1, \dots, \widehat{\g}_L]$ be the matrix whose columns are the first $L$ estimated eigenvectors of interest and $\widehat{\bm{\Gamma}}_{L,(i)} = [\widehat{\g}_{1,(i)}, \dots, \widehat{\g}_{L,(i)}]$ be equivalent to $\widehat{\bm{\Gamma}}_L$ but where the eigenvectors are estimated without the $i^{th}$ observation. Then, $\widehat{\bm{P}}_{L,(i)} = \widehat{\bm{\Gamma}}_{L,(i)} \widehat{\bm{\Gamma}}^\top_{L,(i)}$ is the projection matrix onto the subspace spanned by the columns of $\widehat{\bm{\Gamma}}_{L,(i)}$. B\'{e}nass\'{e}ni's sample influence function is given by, 
\begin{equation}\label{sifBenas}
    \text{SIF}_{\text{B}}(\rho, F_n; \mathbf{x}_i) = (n-1) \left[ \rho \left(\widehat{\bm{\Gamma}}_L, \widehat{\bm{\Gamma}}_{L,(i)} \right) - 1 \right]
\end{equation}
which is based on $ \rho \left(\widehat{\bm{\Gamma}}_L, \widehat{\bm{\Gamma}}_{L,(i)} \right) = 1 - \dfrac{1}{L} \displaystyle\sum_{l = 1}^{L} \norm{(\bm{I} - \widehat{\bm{P}}_{L,(i)}) \widehat{\g}_l}$, a measure concerning the sine of the angle between each $\widehat{\g}_l$ and its projection $\widehat{\bm{P}}_{L,(i)} \widehat{\g}_l$. Then, the associated empirical influence function is given by,
\begin{equation}\label{eifBenas}
        \text{EIF}_{\text{B}}(\rho, F_n; \mathbf{x}_i) = -\dfrac{1}{L} \sum_{l = 1}^{L} \bigg\{ \sum_{k = L+1}^{p} \dfrac{\widehat{\omega}_{li}^2 \widehat{\omega}_{ki}^2}{(\widehat{\lambda}_l - \widehat{\lambda}_k)^2}\bigg\}^{1/2}
\end{equation}
where $\widehat{\omega}_{li} = \widehat{\g}_l^\top (\mathbf{x}_i - \overline{\mathbf{x}} )$ and $\overline{\mathbf{x}}$ is the mean of $\X_n$.

The second influence diagnostic we consider is based on the average squared canonical correlations between, $ \X_n\widehat{\bm{\Gamma}}_L$ and $\X_n \widehat{\bm{\Gamma}}_{L,(i)}$. Here,  $ \X_n\widehat{\bm{\Gamma}}_L = [\widehat{\bm{\alpha}}_1, \dots, \widehat{\bm{\alpha}}_L]$ denotes the subset of the first $L$ SPCs of interest and $\X_n\widehat{\bm{\Gamma}}_{L,(i)} = [\widehat{\bm{\alpha}}_{1,(i)}, \dots, \widehat{\bm{\alpha}}_{L,(i)}]$ is equivalent to $\X_n\widehat{\bm{\Gamma}}_L$ but with the corresponding eigenvectors estimated without the $i^{th}$ observation. Based on this measure \cite{Connie2011} provided the following sample influence function, 
\begin{equation}\label{sifConnie}
    \text{SCI}_i = (n-1)^2 [1 - \overline{r^2}]    
\end{equation}
where $\overline{r^2}$ is the average of the squared canonical correlations between the subset of the SPCs with and without the $i^{th}$ observation. The corresponding empirical influence function is given by, 
\begin{equation}\label{eifConnie}
    \text{SCIA}\big( \overline{r^2}, F_n; \mathbf{x}_i \big) = \dfrac{1}{L} \sum_{l = 1}^L \sum_{k=L+1}^p \dfrac{\widehat{\lambda}_k}{\widehat{\lambda}_l} \dfrac{\widehat{\omega}_{li}^2 \widehat{\omega}_{ki}^2}{\big(\widehat{\lambda}_l - \widehat{\lambda}_k \big)^2}.
\end{equation}
Note here that there is a difference in sign between \eqref{sifBenas} and \eqref{sifConnie} and between \eqref{eifBenas} and \eqref{eifConnie} which is insignificant since it is the magnitude of the measures that is important.

The sample influence functions calculate the true influence of each observation in a data set. However, they can be computationally expensive, especially with big high-dimensional data sets, since they require $n + 1$ eigen-decompositions, at $F_n, F_{n,(1)}, \dots, F_{n,(n)}$. Empirical influence functions, on the contrary, provide approximations to the sample influence and only require one eigen-analysis.  The empirical influence function is derived by replacing the population parameters of the theoretical influence function by the corresponding sample estimates. Therefore, they can be used in practice for their efficiency (since estimation is performed only once) given that they can approximate the sample influence well. For sufficiently large $n$, for example, we have that EIF $\approx$ SIF and SCIA $\approx$ SCI. For more information on influence functions for PCA see, for example, \cite{Benasseni1990SensitivityCF, Prendergast2008, Connie2011}. 

\section{Identifying switching of eigenvalues}\label{sec:Switch}

In this section, we demonstrate the problem of switching through the following motivating example. Then, we show how influence diagnostics can be used to approximate the eigenvalues following the removal of the $i^{th}$ observation.

\subsection{Motivating example: Fatty Acids}\label{subsec:MotivEx}

We consider the Fatty Acids data set \citep{BRODNJAKVONCINA200531} which contains 7 measurements of fatty acid concentrations for 96 commercial oils. The data are used for the classification of vegetable oils into 7 groups: pumpkin (labeled A), sunflower (B), peanut (C), olive (D), soybean (E), rapeseed (F) and corn (G). The data set is freely available through the \texttt{caret} package in R. We now take observation 57 as an example of an observation that causes switching upon removal. When observation 57 is removed from the data, the estimated eigenvalues given by PCA are, 
              
\[ 452.747 \quad  9.850  \quad  9.545  \quad   0.647  \quad  0.369  \quad   0.059 \quad  0.036  .\]

In this case, the estimated eigenvalues are given in descending order and there is no evidence to indicate if switching has occurred. However, simple approximations to the eigenvalues can be found (detailed in Section \ref{subsec:Approx}) that clearly show the disruption in the order of the  eigenvalues. The approximated eigenvalues when considering the removal of the 57$^{th}$ observation are,

\[ 452.727  \quad \bm{ 9.599  \quad 9.816 } \quad   0.647   \quad    0.369  \quad   0.059 \quad   0.036.\]

Here, the approximations reveal switching between the second and third eigenvalues, since we can clearly observe that the third eigenvalue is now greater than the second eigenvalue, $\widehat{\lambda}_2 < \widehat{\lambda}_3$. Identifying the disruption in the order of the eigenvalues is important because the removal of such observations also disrupt the order of the corresponding eigenvectors which in turn affect the corresponding SPCs.
This example will be revisited in Section \ref{sec:Examples} to show one of the main benefits of knowing when switching has occurred.

\subsection{Approximating the eigenvalues}\label{subsec:Approx}

The influence function \cite[IF,][]{hampel1974influence} is used to assess the robustness of estimators to a small amount of contamination in the population or, in the sample setting, sensitivity to the removal of a single observation. Consider a distribution function $F$, and let $F_\epsilon = (1-\epsilon)F+\epsilon \Delta_{\x}$ where $0<\epsilon << 1$ and $\Delta_{\x}$ is the distribution function that puts all of its mass at the contaminant point $\x$.  Then the influence function for an estimator that has statistical functional $T$, is
\begin{equation}
    \IF(T, F; \x)=\lim_{\epsilon\downarrow 0}\frac{T(F_\epsilon)-T(F)}{\epsilon}=\left.\frac{\partial}{\partial\epsilon}T(F_\epsilon)\right|_{\epsilon =0}.
\end{equation}

In the context of this work, it is the sample versions of influence functions that are relevant and these will be detailed more later.  For readers interested in influence function theory and examples, refer to, e.g. \cite{hampel1974influence, Devlin75} and \cite{hampel2011robust}. 

There is nothing specifically new in the lemma presented below concerning influence functions for an eigenvalue, other than that some conditions have been relaxed for $W$ as to what is commonly found in the literature.  For example, \cite{croux2000principal} provide the influence function for the $j^{th}$ eigenvalue as part of their Lemma 3 where $W$ is also positive definite, rather than just symmetric.  For completeness, a sketch proof is provided in the Appendix and mirrors or is similar to derivations found in the literature.  In this multivariate setting, $\mathbf{x}_0\in \mathbb{R}^p$ denotes the $p$-dimensional contaminant vector.

\begin{lemma}\label{lemma:IFlj}
Let $W$ be the statistical functional for a symmetric $p$-dimensional matrix estimator and $F$ a $p$-dimensional distribution function where $W(F)=\mathbf{W}$.  Let $l_j$ and $e_j$ be the statistical functionals for the $j^{th}$ eigenvalue and eigenvector where, at $F$, $W(F)e_j(F)=l_j(F)e_j(F)=\mathbf{W}\bm{\eta}_j=\lambda_j\bm{\eta}_j$ and where $\lambda_j$ is a unique eigenvalue.  Then, 
\begin{equation*}
    \IF (l_j ,F; \x)= \bm{\eta}_j^\top \IF(W, F; \x) \bm{\eta}_j.
\end{equation*}
\end{lemma}

Consider the following results which can be used to approximate eigenvalues following the removal of an observation.  A discussion as to how this result arises will follow.

\begin{result}\label{result:lji}
Suppose that $\widehat{\mathbf{W}}$ is a symmetric matrix estimate where estimation has been carried out on $n$ observations.  Let $\widehat{\bm{\eta}}_j$ be the estimated $j^{th}$ eigenvector corresponding to the $j^{th}$ ordered eigenvalue $\widehat{\lambda}_j$.  Also let $\widehat{\mathbf{W}}_{(i)}$ and $\widehat{\lambda}_{j,(i)}$ denote the corresponding estimates when the $i^{th}$ observation is removed.  Then, under some mild conditions,
$$\widehat{\lambda}_{j,(i)} \approx \widehat{\lambda}_j - \widehat{\bm{\eta}}_j^\top \left(\widehat{\mathbf{W}}_{(i)} -\widehat{\mathbf{W}} \right)\widehat{\bm{\eta}}_j= \widehat{\bm{\eta}}_j^\top\widehat{\mathbf{W}}_{(i)}\widehat{\bm{\eta}}_j. $$
\end{result}

Result \ref{result:lji} has previously been used to approximate eigenvalues for principal component analysis after removal of a set of observations \citep[e.g.][]{benasseni2018correction}.  Now we explain how it arises for eigenvalue estimates from any symmetric matrix estimator using the idea of the hybrid influence function (HIF) introduced by \cite{prendergast2007implications} and used by \cite{shakerthesis} to improve empirical influence function (EIF) approximations of the sample influence function (SIF).  Using the notations above but noting that the idea of sample influence functions generalises to other estimators, the SIF for the $j^{th}$ eigenvalue estimate is 
\begin{equation}\label{sif:lj}
    \SIF(l_{j}, F_n;\mathbf{x}_i)= -(n-1) \left[ \widehat{\lambda}_{j,(i)} - \widehat{\lambda}_{j} \right]
\end{equation}
where $F_n$ denotes the empirical distribution function and $\mathbf{x}_i$ the $i^{th}$ observation.  Hence, $ \SIF(l_{j}, F_n; \mathbf{x}_i)$ is a measure of influence associated with the removal of the $i^{th}$ observation.  If the IF exists and is known in a closed-form expression, then one could use the EIF defined to be $\EIF(l_j, F_n; \mathbf{x}_i)=\IF(l_j,F_n;\mathbf{x}_i)$ so that the parameters in $\IF(l_j, F; \x)$ are replaced with their estimates from $F_n$ and $\mathbf{x}_i$ becomes the contaminant.  We then have that $\EIF(l_j,F_n;\mathbf{x}_i)$ is an approximation for $\SIF(l_j,F_n;\mathbf{x}_i)$.  To make this clear we consider an example that will also be useful later.

\begin{example}\label{ex:C}
Let $C$ denote the functional for the covariance matrix estimator with $\bm{\Sigma}=C(F)=\int (\bm{X}-\bm{\mu})(\bm{X}-\bm{\mu})^\top dF$ where $\bm{X}\sim F$ and $\bm{\mu}=E(\bm{X})$ is the mean vector.  Then, consider a sample of $n$ observations given as $\mathbf{x}_1,\ldots,\mathbf{x}_n$ where $\bm{\Sigma}$ is estimated by $\widehat{\bm{\Sigma}}=C(F_n)=n^{-1}\sum^n_{i=1}(\mathbf{x}_i-\overline{\mathbf{x}})(\mathbf{x}_i-\overline{\mathbf{x}})^\top$ where $\overline{\mathbf{x}}$ is the sample mean.  The IF for $C$ is \cite[e.g.][]{critchley1985influence}, $\IF(C,F; \x)=(\x-\bm{\mu})(\x-\bm{\mu})^\top- \Si $ which leads to the EIF for the $i^{th}$ observation as $\EIF(C,F_n;\mathbf{x}_i)=(\mathbf{x}_i-\overline{\mathbf{x}})(\mathbf{x}_i-\overline{\mathbf{x}})^\top-\widehat{\bm{\Sigma}}$ which approximates the SIF  $=(n-1)\left(\widehat{\bm{\Sigma}}_{(i)} - \widehat{\bm{\Sigma}}\right)$.  Unless $n$ is small, this approximation is very close and evidence of this is given in Equations 13 and 14 of \cite{prendergast2007implications}.
\end{example}

In the context of eigenvalues, from Lemma \ref{lemma:IFlj}, the EIF for the $j^{th}$ sample eigenvalue is $\EIF(l_j, F_n;\mathbf{x}_i)=\widehat{\bm{\eta}}_j^\top \EIF(W, F_n;\mathbf{x}_i)\widehat{\bm{\eta}}_j$.  If $W$ was the functional for the covariance matrix estimator considered in Example \ref{ex:C}, then the EIF exists in a closed form and is given as $\EIF(l_j,F;\mathbf{x}_i)=\left[\widehat{\bm{\eta}}_j^\top(\mathbf{x}_i-\overline{\mathbf{x}})\right]^2-\widehat{\lambda}_j$.   However, if the IF, and hence the EIF is unknown, then we could use the SIF in place of the EIF.  This is called the hybrid influence function \citep[HIF,][]{prendergast2007implications, shakerthesis} and in our context of the $j$th eigenvalue we have
\begin{align}
    \HIF(l_j, F_n;\mathbf{x}_i)=& \widehat{\bm{\eta}}_j^\top \SIF(W, F_n;\mathbf{x}_i)\widehat{\bm{\eta}}_j\nonumber\\
    =& -(n-1)\widehat{\bm{\eta}}_j^\top \left(\widehat{\mathbf{W}}_{(i)} - \widehat{\mathbf{W}}\right)\widehat{\bm{\eta}}_j.
\end{align}

Finally, Result \ref{lemma:IFlj} arises due to the fact that $\HIF(l_j, F_n;\mathbf{x}_i)\approx \SIF(l_j, F_n;\mathbf{x}_i)$ and a simple rearrangement, recalling the SIF is given in \eqref{sif:lj}.  Hence, we have a convenient means to approximate our eigenvalues post-removal of an observation for any dimension reduction method, whether or not a closed form solution for the theoretical influence function exists.

\section{Example: Fatty Acids Data}\label{sec:Examples}

\begin{figure}
    \centering
    \includegraphics[width = \textwidth]{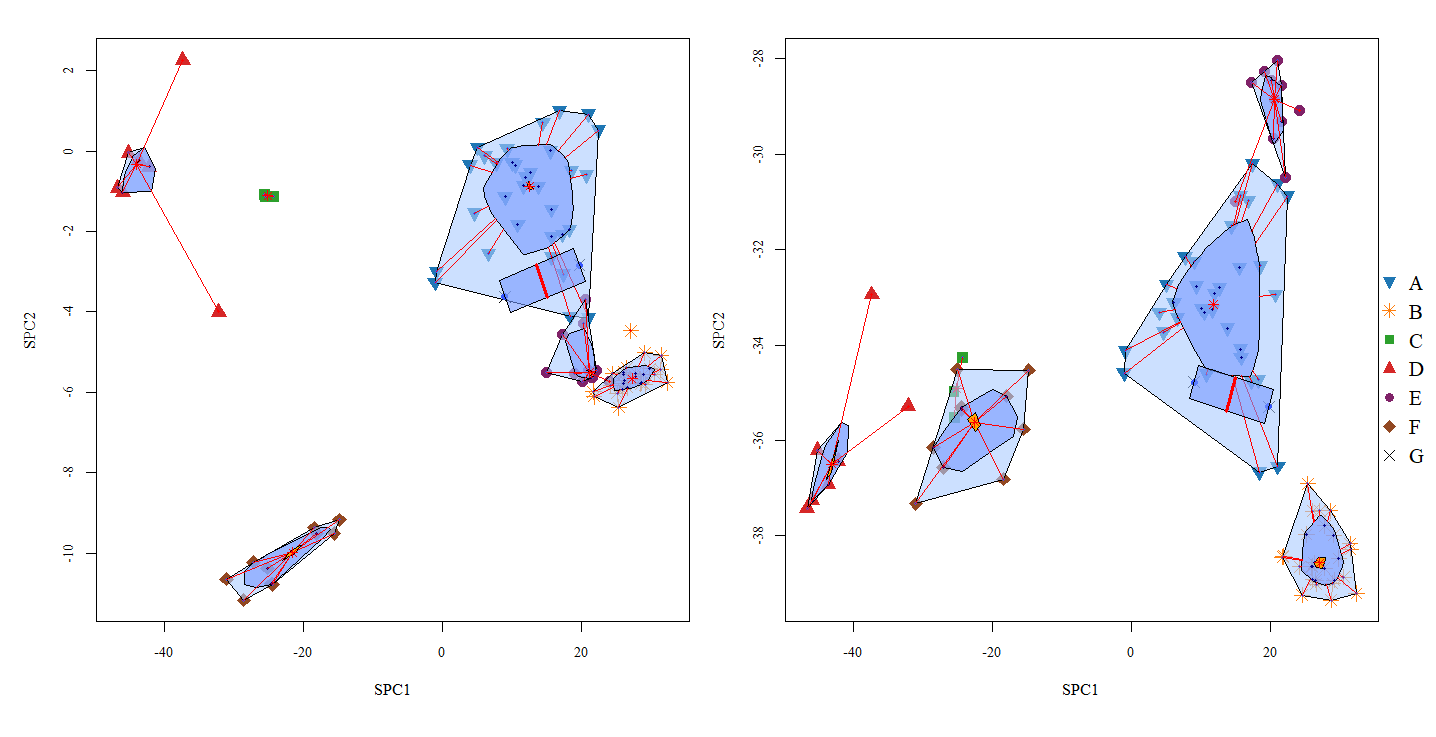}
    \caption{Biplots of the Fatty Acids data set with all observations (left) and with observation 57 removed from the estimation of the eigenvectors (right). Bagplots overlay the scatter plots to provide an interesting visualization of the groups in the data.}
    \label{fig:OilsBagplot}
\end{figure}

In this section, we return to the Fatty Acids data set from the motivating example in section \ref{subsec:MotivEx}. We previously found that observation 57 results in switching between $\widehat{\lambda}_2$ and $\widehat{\lambda}_3$. Similarly, using the approximations introduced in Result \ref{result:lji}, observations 42, 58, 59, 60, 91 and 93 also result in switching between the second and third eigenvalues.  Note that no switching was observed between $\widehat{\lambda}_1$ and $\widehat{\lambda}_2$, and $\widehat{\lambda}_3$ and $\widehat{\lambda}_4$. As previously mentioned, the switching between the eigenvalues also affects the corresponding eigenvectors and therefore the SPCs that they form. 
It follows that when, for example, observation 57 is removed, the estimated eigenvectors that correspond to $\widehat{\lambda}_2$ and $\widehat{\lambda}_3$ are quite different pre- and post-removal.
This is evident in Figure \ref{fig:OilsBagplot} which shows the biplots, overlayed with bagplots to emphasise groupings, produced with and without the $57^{th}$ observation. It is clear that the plots provide vastly different visualizations due to the removal of a single observation from the data. This indicates the large influence of observation 57 on the SPCs.  The effect of such an observation is not the same when we consider the 3D plot of the first three SPCs. If both of the eigenvectors involved in switching are considered, then these observations are no longer highly influential on the retained subset of SPCs. Therefore, one reason it can be important to know where switching has occurred is for choosing the number of components ($L$) to retain, so it can be ensured that no switching occurs between $L$ and $L+1$. For the Fatty Acids data set, the obvious choice is $L = 3$. 

\begin{figure}
    \centering
    \includegraphics[width = \textwidth, page = 1]{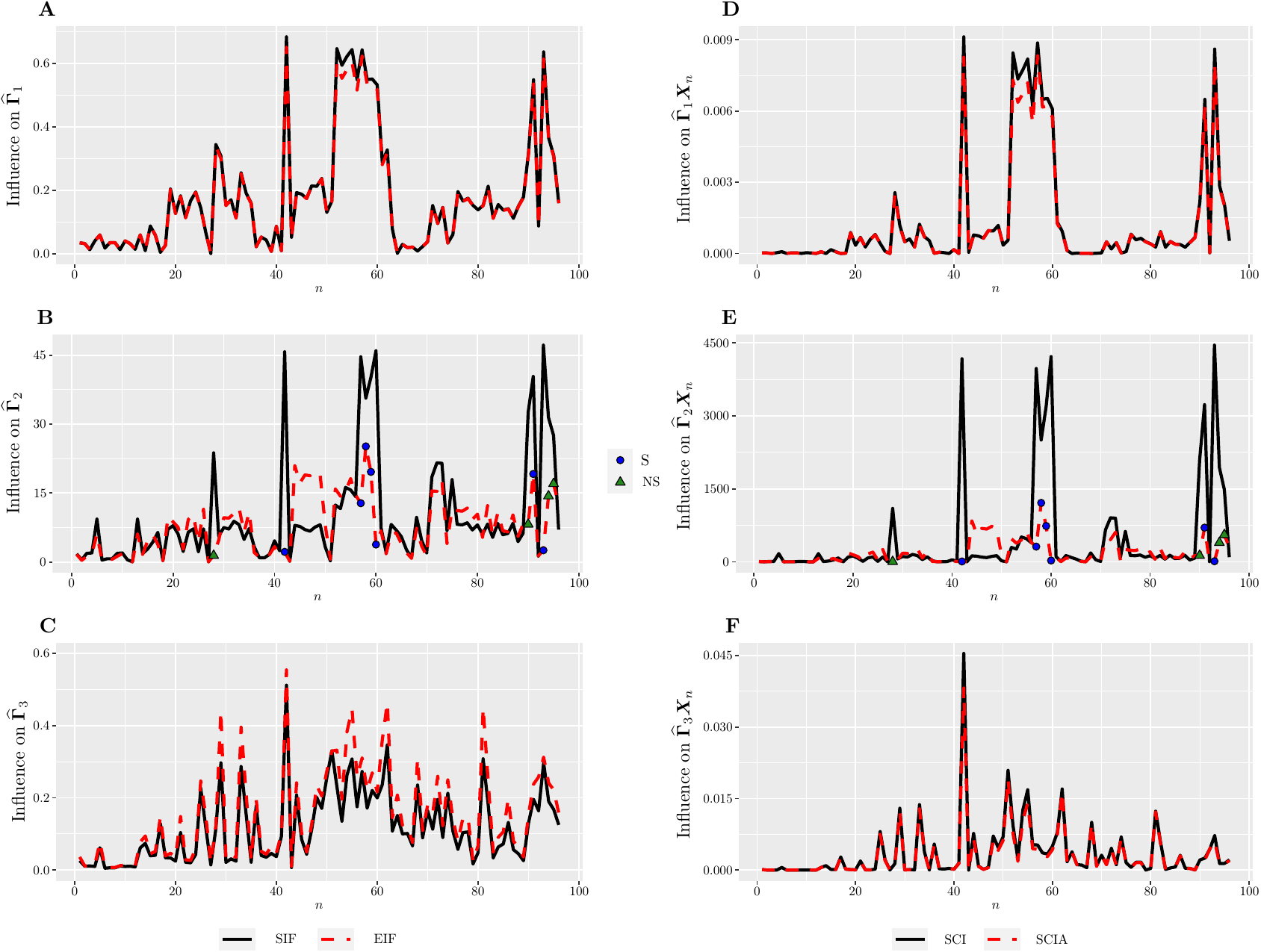}
    \caption{A, B and C are plots comparing the SIF ($-$) in \eqref{sifBenas} and EIF (- -) in \eqref{eifBenas} for the Fatty Acids data with $L = 1, 2$ and 3, respectively. D, E and F are plots comparing the SCI ($-$) in \eqref{sifConnie} and SCIA (- -) in \eqref{eifConnie} for the Fatty Acids data with $L = 1, 2$ and 3, respectively. The blue circle and green triangle points, in plots B and E, are denoted by ``S" and ``NS" to indicate the observations that were identified to cause switching and the ones that did not result in switching, respectively.  }
    \label{fig:OilVectorIFs}
\end{figure}

Now consider the influence measures introduced in section \ref{sec:PCA_IFs} and the associated notation. From the above, we expect the observations that cause switching to be highly influential on $\widehat{\bm{\Gamma}}_2$ and $ \X_n\widehat{\bm{\Gamma}}_2$. We also expect them to have relatively small influence on $\widehat{\bm{\Gamma}}_1$, $\X_n\widehat{\bm{\Gamma}}_1$, $\widehat{\bm{\Gamma}}_3$ and $\X_n\widehat{\bm{\Gamma}}_3$. Indeed, Figure \ref{fig:OilVectorIFs}, supports these claims. Plots A, B and C, in Figure \ref{fig:OilVectorIFs}, compare the values of the SIF$_\text{B}$ ($-$) and EIF$_\text{B}$ (- -) measures (given in \eqref{sifBenas} and \eqref{eifBenas}) for this data, for $L = 1, 2$ and 3, respectively. Plots D, E and F compare the values of the SCI ($-$) and SCIA (- -) measures (given in \eqref{sifConnie} and \eqref{eifConnie}) for $L = 1, 2$ and 3, respectively. 

In Plot A and D, the EIF$_\text{B}$ and SCIA provide good approximations to the SIF$_\text{B}$ and SCI measures, respectively, and influence values are relatively small (less than 1) as expected. The same is true for Plot C and F.
The EIF$_\text{B}$ and SCIA, however, in plot B and E, provide poor approximations to the SIF$_\text{B}$ and SCI for the observations found to cause switching. The EIF$_\text{B}$ and SCIA values of those observations are highlighted by blue circle points. As previously mentioned, the SIF$_\text{B}$ and SCI find the true influence of the observations in the data. The EIF$_\text{B}$ and SCIA, however, utilize the original estimates (found by considering all observations in the data) to approximate the true influence of each observation. Since there is no evidence of switching in those estimates, empirical measures underestimate the influence of switching observations, sometimes to the extent of approximating a zero influence value (for example, observation 42 in Plots B and E of Figure \ref{fig:OilVectorIFs}).  Empirical influence functions are very useful in practice due to their efficiency and it is of great importance that they provide good approximations to the associated sample influence diagnostics. However, when switching occurs empirical influence diagnostics can be misleading. Nevertheless, since we can now identify switching observations, we can retain the efficiency of the EIF$_\text{B}$ by replacing the influence of just those observations with the SIF$_\text{B}$. The same applies for the SCI and SCIA measures or any other influence diagnostics. Hence, we preserve the efficiency of empirical measures since the computational cost of calculating the sample influence of just a few observations is very small compared to finding the sample influence for all observations in the data.

Note here that to find the observations that result in switching we are searching for all observations where, for example, $\widehat{\lambda}_{2,(i)} < \widehat{\lambda}_{3,(i)}$. This criterion is used for each pair of consecutive approximated eigenvalues given by Result \ref{result:lji}. Besides the observations identified from the above, Plots B and E also have some other observations for which the EIF$_\text{B}$ and SCIA do not agree with the SIF$_\text{B}$ and SCI respectively. These observations (28, 90, 94 and 95) are marked by green triangle points and denoted by ``NS". We found that these observations have $\widehat{\lambda}_{2,(i)}$ and $\widehat{\lambda}_{3,(i)}$ values very close together. Therefore, another criterion for finding these additional observations is for example, $| \widehat{\lambda}_{2,(i)} - \widehat{\lambda}_{3,(i)}| < 0.1$ (which also finds some of the switching observations). Keep in mind that each data set is different and the value 0.1 will not necessarily be suited to other examples. Since some of the ``NS" observations are quite influential, then we can replace their empirical influence values with their true influences as we proposed for the switching observations. 

\begin{figure}
    \centering
    \includegraphics[width = \textwidth, keepaspectratio]{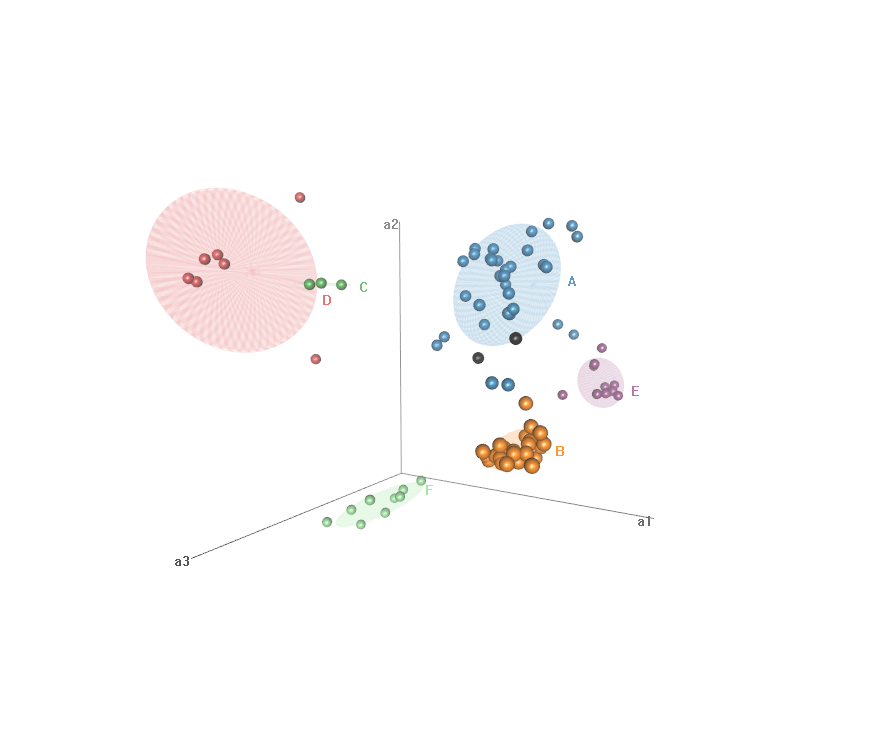}
    \vspace{-3cm}
    \caption{3D scatter plot of the first 3 SPCs for the Fatty Acids data set.}
    \label{fig:Oils3D}
\end{figure}

Finally, Figure \ref{fig:Oils3D} shows the 3-dimensional plot of SPC1 versus SPC2 versus SPC3. It is evident, that the 3D plot provides a better visualization of the oil groups compared to the 2D plots in Figure \ref{fig:OilsBagplot}. This supports the choice of L = 3 based on the the presence of switching between $\widehat{\lambda}_2$ and $\widehat{\lambda}_3$. 

\section{Recommendations for switching}\label{sec:Suggestions}

Identifying when switching occurs and which observations are responsible gives us the opportunity to make informed decision and adjust our methodologies. Here we discuss a few recommendations of potential actions one can take when switching affects the results of the analysis. 

\textbf{SIF replacement}

Empirical influence diagnostics are very useful in practice, especially when we want to perform the analysis of big high-dimensional data sets quickly. However, the Fatty Acids example in Section \ref{sec:Examples} showed that empirical measures fail to approximate the influence of observations that result in switching, which are some of the most highly influential observations in the data. The ability to identify these observations provides a great advantage in this case. Replacing the approximated influences of such observations by their sample influences is an easy and simple solution to this problem. \cite{liwaisuenthesis} argues that if the SIF replacement is required for several observations, the computational efficiency gains of the empirical approximations might be lost. However, even though the computation of the SIF can be costly, it does not compare with the time-consumption of having to calculate the SIF for all observations in a high-dimensional data set. Therefore, SIF replacement is the most effective and efficient way to obtain accurate influence values when switching occurs. 

\textbf{Number of lower-dimensional variables to retain}

While in PCA we use the term principal components, in other methods, the new set of lower dimensional variables are referred to as effective dimension reducing (e.d.r) predictors. For simplicity we will use PCA to present the following recommendation. Keep in mind however that what follows directly applies to the choice of e.d.r predictors in other methods. Note here that the true number of SPCs is unknown in practice. A plethora of tests have been formulated that can be used to decide on how many SPCs should be retained. However, if a test indicates $L$ components and switching occurs between the $L^{th}$ and $L+1$ eigenvalues, then either $L-1$ or $L+1$ SPCs should be retained instead. The number of SPCs to retain should not be sensitive to the removal of any particular observation from the data. 

\textbf{Manually re-ordering the eigenvalues and eigenvectors}  

\cite{critchley1985influence} proposed that the re-ordering of the eigenvalues and eigenvectors of observations that result in switching is sufficient. However, \cite{liwaisuenthesis} showed through an example \citep[see, Example 4.2.2,][]{liwaisuenthesis} that re-ordering the eigenvectors does not necessarily provide a solution to the problem of switching. 
Furthermore, if several observations result in switching, then manually re-ordering the eigenvalues and eigenvectors can quickly become cumbersome. It is safe to say that this would not be an elegant way to deal with switching.

\section{Discussion}\label{sec:Discuss}

In this article, we utilized influence diagnostics to provide simple approximations to the eigenvalues of any symmetric matrix estimate in the dimension reduction setting. These approximations are able detect observations whose removal results in switching between consecutive eigenvalues. Specifically, we considered the sensitivity of eigenvalues of the covariance matrix (i.e. PCA) to demonstrate the problems brought on by switching. 
For example, the inability of the EIF to approximate how influential such observations are on the estimators of interest. We also showed, through a real-world example, that switching should be considered when choosing the number of SPCs to retain. Finally, we provided a few recommendations that are  beneficial when switching is detected in practise.

Another possible action one might suggest for dealing with switching would be to delete these observations from the data. However, our explorations showed that this is not as simple as it sounds since deleting those observations might result in other observations causing switching. This is not always true of course and in some cases it might prove to be beneficial.

\appendix

\section{Proof of Lemma \ref{lemma:IFlj}}

Using the product rule on $ l_j(\Fe) = e_j^\top(\Fe)  W(\Fe) e_j(\Fe) $, we get, 
\begin{equation}
    \dfrac{\partial}{\partial \e} l_j(\Fe)  = \left[\dfrac{\partial}{\partial \e} e_j^\top(\Fe)\right] W(\Fe) e_j(\Fe) +   e_j^\top(\Fe) \left[\dfrac{\partial}{\partial \e} W(\Fe)\right] e_j(\Fe) + e_j^\top(\Fe) W(\Fe) \left[\dfrac{\partial}{\partial \e} e_j(\Fe)\right]
\end{equation}

Noting that, $W(\Fe)e_j(\Fe) =  l_j(\Fe) e_j^\top(\Fe)$, we have, 
\begin{align}\label{IFl}
    \dfrac{\partial}{\partial \e} l_j(\Fe) &= \left[ \dfrac{\partial}{\partial \e}e_j^\top(\Fe)\right]  e_j(\Fe) l_j(\Fe) e_j^\top(\Fe) e_j(\Fe) + e_j^\top(\Fe) \left[\dfrac{\partial}{\partial \e} W(\Fe)\right] e_j(\Fe) \nonumber \\ 
    &\quad +  e_j^\top(\Fe) e_j(\Fe) l_j(\Fe) e_j^\top(\Fe) \left[\dfrac{\partial}{\partial \e} e_j(\Fe)\right] 
\end{align}

Now, the eigenvectors are orthonormal, so that, 
\begin{equation}\label{orthonormality}
    e_j^\top(\Fe) e_j(\Fe) = 1.
\end{equation}
Using the Product Rule, it follows  that,
\begin{equation}\label{orthnormalConsequence}
   \left[ \dfrac{\partial}{\partial \e} e_j^\top(\Fe) \right] e_j(\Fe) + e_j^\top(\Fe) \left[ \dfrac{\partial}{\partial \e} e_j(\Fe) \right] = 0
\end{equation}

Then, from \eqref{orthonormality} and \eqref{orthnormalConsequence}, and since $l_j(\Fe)$ is a scalar, \eqref{IFl} becomes, 
\begin{align}
    \dfrac{\partial}{\partial \e} l_j(\Fe) &= \Bigg(\left[ \dfrac{\partial}{\partial \e}e_j^\top(\Fe)\right]  e_j(\Fe) + e_j^\top(\Fe) \left[\dfrac{\partial}{\partial \e} e_j(\Fe)\right] \Bigg)l_j(\Fe) + e_j^\top(\Fe) \left[\dfrac{\partial}{\partial \e} W(\Fe)\right] e_j(\Fe) \nonumber\\
    &= e_j^\top(\Fe) \left[\dfrac{\partial}{\partial \e} W(\Fe)\right] e_j(\Fe)
\end{align}

The proof is complete by setting $\e$ to 0.

\bibliographystyle{authordate4}
\bibliography{refs.bib}

\end{document}